# Raman lasing and soliton modelocking in lithium-niobate microresonators

Mengjie Yu[1, *], Yoshitomo Okawachi[2, *], Rebecca Cheng[1], Cheng Wang[3], Mian Zhang[4], Alexander L. Gaeta[2], & Marko Lončar[1]

[1]Department John A. Paulson School of Engineering and Applied Sciences, Harvard University, Cambridge, MA 02138

[1]Department of Applied Physics and Applied Mathematics, Columbia University, New York, NY 10027

[2]School of Electrical and Computer Engineering, Cornell University, Ithaca, NY 14853

[3]Department of Electronic Engineering & State Key Laboratory of Terahertz and Millimeter Waves, City University of Hong Kong, Kowloon, Hong Kong, China

[4]HyperLight Corporation, 501 Massachusetts Avenue, Cambridge, MA 02139

[4]These authors contributed equally.

## Abstract

The recent advancement in lithium niobate on insulator (LNOI) technology is revolutionizing the optoelectronic industry as devices of higher performance, lower power consumption, and smaller footprint can be realized due to the high optical confinement in the structures. The LNOI platform offers both large $\chi^{(2)}$ and $\chi^{(3)}$ nonlinearities along with the power of dispersion engineering, enabling brand new nonlinear photonic devices and applications towards the next generation of integrated photonic circuits. However, the Raman scattering, one of the most important nonlinear phenomena, have not been extensively studied, neither was its influences in dispersion-engineered LNOI nano-devices. In this work, we characterize the Raman radiation spectra in a monolithic lithium niobate (LN) microresonator via selective excitation of Raman-active phonon modes. Remarkably, the dominant mode for Raman oscillation is observed in the backward direction for a continuous-wave pump threshold power of 20 mW with a reportedly highest differential quantum efficiency of 46 %. In addition, we explore the effects of Raman scattering on Kerr optical frequency combs generation. We achieve, for the first time, soliton modelocking on a X-cut LNOI chip through sufficient suppression of the Raman effect via cavity geometry control. Our analysis of the Raman effect provides guidance for the development of future chip-based photonic devices on the LNOI platform.

**Introduction**

The monolithic lithium niobate on insulator (LNOI) platform has attracted significant interest for realization of next-generation nonlinear photonic devices and observation of new nonlinear dynamics due to its large $\chi^{(2)}$ ($r_{33} = 3 \times 10^{-11}$ m/V) and $\chi^{(3)}$ nonlinearities ($n_2 = 1.8 \times 10^{-19}$ m$^2$/W) [1-16]. The LNOI platform is opening new opportunities for large-scale integration of optical and electronic devices on a single chip, as it combines the material properties of lithium niobate with the integration power of nano-photonics. By leveraging the recent advances in the fabrication of ultra-low loss lithium niobate (LN) nanowaveguides and microring resonators [1], researchers have demonstrated Kerr optical frequency combs (OFC's) [2-4], broadband electro-optic combs [5], highly efficient second harmonic generation [6-8], and multiple-octave-spanning supercontinuum generation (SCG) [8,9]. LN is known as a Raman-active crystalline material with several strong vibrational phonon branches in different polarization configurations [17-23]. There has been evidence of Raman scattering in LN disks or whispering gallery resonators [24,25]. However, such platforms are fabricated by the mechanical polishing method [26] and their group velocity dispersion (GVD), which is critical for nonlinear parametric processes, cannot be engineered. The Raman effect in integrated photonic devices not only enables Raman lasers for generating new frequencies at low optical powers [27-34], but also can lead to nontrivial nonlinear interactions through tailoring the dispersion properties, such as interplay between Raman and both $\chi^{(2)}$ and $\chi^{(3)}$ effects, impacting Kerr comb formation, electro-optic comb formation, and supercontinuum generation [35-41]. Recent work by Hansson, *et. al.*, has shown that aligning the cavity free-spectral range (FSR) to the Raman gain can allow for the generation of an octave-spanning Raman frequency comb [42]. Among them, the LN photonic circuit is particularly appealing to microresonator-based Kerr frequency comb generation since the presence of large second order nonlinearity $\chi^{(2)}$ offers advantageous functionality for fully on-chip optical clock and metrology, a key element missing from the current mature silicon nitride or silica technologies. Due to the large Raman effect in crystalline material, strong interplay between Raman scattering and four-wave mixing (FWM) has been observed in materials such as diamond and silicon [36], and strategies have been proposed to suppress these interactions by controlling the FSR [36,39,43]. However, Raman scattering and its influence on soliton modelocking remain largely unexplored in dispersion-engineered monolithic LNOI devices.

In this paper, we demonstrate multiwavelength Raman lasing in an X-cut high-$Q$ LN microresonator with Raman frequency shifts of 250 cm$^{-1}$, 628 cm$^{-1}$, and 875 cm$^{-1}$ via pumping with transverse-electric (TE) polarization, and of 238 cm$^{-1}$ with traverse-magnetic (TM) polarization. The dominant Raman oscillation occurs in the backward direction with respect to the pump, and the backward Raman gain coefficient is measured to be 1.3 cm/GW for the 250 cm$^{-1}$ Raman shifted line for TE polarization, and 0.07 cm/GW for a 238 cm$^{-1}$ Raman shifted line for TM polarization. In both cases, a 1.5 μm pump is used to excite the sample. To our knowledge, this is the first characterization of multiple Raman lasing on a monolithic LN chip. In addition, we investigate the effects of the Raman process on Kerr comb generation and soliton modelocking for both polarizations and show that the Raman effect can be controlled to enable modelocked Kerr comb formation for TM polarization.

# Results

LN is a uniaxial material with its crystal axis along z direction as shown in Fig. 1(a). The LN devices are fabricated on an X-cut thin-film wafer where the x-axis is normal to the wafer plane. LN has two Raman-active phonon symmetries: the A symmetry polarized along z-axis and the E symmetry polarized in the degenerate x-y plane [19,21] due to the atomic vibration. Furthermore, both transverse (TO) and longitudinal (LO) optical phonon modes exist. X-cut wafer allows for accessing both TO and LO modes in both symmetries. The selection rules of Raman scattering depend on the wavevectors and polarization of the pump and Stokes fields [17-20]. Two different cavity geometries are used in our experiments [Fig. 1(b,c)], where TE polarized light is mostly parallel with the crystal axis in the racetrack geometry and TM polarized light is parallel with non-polar axis (x-axis).

## Characterization of Raman scattering

The experimental setup for Raman characterization is shown in Fig. 1(d). We inject an amplified continuous-wave (CW) pump laser centered at 1560-nm into a monolithically integrated LN racetrack microresonator [Fig. 1(b)]. The device is cladded with silicon oxide with a top waveguide width of 1.2 μm and an etch depth of 450 nm on a 800-nm thick LN thin film. The racetrack design allows for two long straight waveguide regions to maximize the interaction with the TO optical phonon mode for TE polarization. The FSR of the microresonator is 30 GHz which is within the Raman gain bandwidth [19]. The intrinsic-$Q$ of the resonator is $\sim 1.5 \times 10^6$ for both TE and TM modes (see Supplementary Information). We record the Raman emission spectra in both forward and backward direction using an optical circulator and two optical spectrum analyzers at various pump powers in the bus waveguide. For TE polarization, we observe several Raman oscillations [Fig. 2(a)] with frequency shifts of 250 cm$^{-1}$ (7.5 THz), 628 cm$^{-1}$ (18.8 THz), and 875 cm$^{-1}$ (26.2 THz), corresponding to the optical phonon branches of A(TO)$_1$, A(TO)$_4$, and A(LO)$_4$, respectively. The 1$^{st}$ Stokes (250 cm$^{-1}$) has the lowest pump threshold of 20-mW with a high differential conversion efficiency of 46 % as shown in Fig. 2(b). This is, to our knowledge, the highest quantum conversion efficiency reported in LN material. As the pump power increases, a mini-comb starts to form around the 1$^{st}$ Stokes peak due to the anomalous GVD. This mini-comb prevents the first Stokes line from monotonic increasing in power. In addition, cascaded Raman peaks form around 1691 nm at 170-mW pump power at the forward direction. Notably, strong spectral peaks at 1691 nm are also observed in the backward direction largely due to a FWM process where the dominant 1$^{st}$ Stokes line acts as the pump. 2$^{nd}$ and 3$^{rd}$ Stokes appears as the pump power reaches 200 mW and 400 mW, respectively. The efficiency of the Raman effect is higher in the backward direction which is phase-matched [44]. This is particularly true for microscale waveguides that feature a non-negligible longitudinal electric field component [45]. This asymmetric gain can also be attributed to strong polaritonic effects which affect phase-matching conditions in the forward direction [24,46]. We estimate the Raman gain $g_R$ of the 1$^{st}$ Stokes line to be 1.3 cm/GW based on [33]. For the 2$^{nd}$ and 3$^{rd}$ Stokes peaks, we are unable to extract the Raman gain due to the presence of Kerr and other Raman processes influencing the pump power. Previously, Raman gain in bulk LN of the corresponding mode was reported to be 12.5 cm/GW at 488 nm by Bache [23], 8.9 cm/GW at 694 nm by Boyd [21] and 5 cm/GW at 1047 nm by Johnson & Chunaev et al. [22], in good agreement with our measurement based on the relation $g_R \propto (\lambda_p \lambda_s)^{-1}$, where $\lambda_p$ is the pump wavelength and $\lambda_s$ is the Stokes wavelength [29]. Similarly, we characterize the devices using TM pump, where the light polarization is along x-axis. As shown in Fig. 3, only one Raman oscillation is observed with a frequency shift of 238 cm$^{-1}$ [E(TO)$_3$] at a threshold power of 340 mW, which corresponds to a Raman gain of 0.07 cm/GW.

**Kerr comb generation and modelocking**

Next, we investigate the effects of Raman scattering on Kerr comb formation for both TE and TM polarizations. Modelocked Kerr frequency combs is particularly attractive on a LN chip for optical metrology since the combination of large $\chi^{(2)}$ and $\chi^{(3)}$ nonlinearities could enable direct on-chip self-referencing without external amplifiers or periodically poled LN crystal. In order to achieve soliton modelocking in the presence of strong Raman scattering, a microring with a smaller radius is preferred to favor the broadband parametric gain (0.15 cm/GW) [36]. Here, we pump an air-cladded LN microring resonator with an 80 μm radius, which corresponds to a 250 GHz FSR [Fig. 1(b)]. The pump power in the bus waveguide is 400 mW for both polarizations. The LN devices here are air-clad with a top waveguide width of 1.3 μm and an etch depth of 350 nm on an X-cut 600-nm thick LN thin film. The cross section is engineered to allow for anomalous GVD for both polarizations (see Supplementary Information). We measure a loaded-$Q$ of >1.5 × $10^6$ for both TE and TM modes. Figure 4 shows the Kerr comb generation dynamics for the TE polarization. We measure the generated spectrum as the pump is tuned into a cavity resonance. As power builds up in the cavity, we observe strong Raman peaks that correspond to the phonon branch of A(TO)$_4$ and A(LO)$_4$ [Fig. 4(a), top] while the A(TO)$_1$ mode is successfully suppressed. With further pump detuning, we observe the formation of primary sidebands, due to the parametric gain, for Kerr comb formation [(Fig. 4(a), middle] and mini-comb formation around the primary sidebands [Fig. 4(a), bottom] [47]. The RF spectrum corresponding the bottom state in Fig. 4(a) is shown in 4(b), indicating that the comb is in a high noise state. This is largely due to the strong Raman effect that competes with the FWM interaction and prevents modelocking [35,36].

Figure 5 shows the comb dynamics for the TM mode. Unlike the TE mode, the Raman effect is much weaker, and we do not observe Raman oscillation at these pump powers. This is attributed to the fact that the Raman gain is less than the Kerr gain at the larger FSR. Figure

5(a) shows the generated spectra for increased red-detuning of the pump. We observe the primary sidebands [State (i)], high-noise [State (ii)], and the multi-soliton state [State (iii)]. The RF spectra [Fig. 5(b)] show the reduction in RF noise from State (ii) to State (iii). Figure 5(c) shows the transmission measurement of the resonator as the pump wavelength is swept through the resonance. The output is optically filtered using a longpass filter with a cut-on wavelength of 1570 nm. Figure 5(d) shows a zoom-in of the dashed-rectangular region in Fig.5(c) that indicates the 'soliton step' representing soliton formation [48]. This indicates that operating in the TM mode using X-cut LN thin film, allows for sufficient suppression of Raman effects enabling modelocked Kerr comb generation. In contrast to Ref. [3] in a Z-cut LN thin film, where the behavior of self-starting or bidirectional tuning is observed as a result of the photorefractive effect, our microresonator is dominated by thermo-optic effect. Moreover, our results do not indicate the occurrence of Raman self-frequency shifting observed by Ref. [3] which often occurs in amorphous material. The difference in dynamics may be attributed to thin film with different crystal orientations. We report the first demonstration of soliton modelocking in X-cut LN microresonators compatible with active electrode control [2,5].

## Discussion

In conclusion, we achieve multiple Raman lasing on a monolithic LN chip, and characterize the distinct Raman processes for different pump polarizations. All the Raman oscillations are dominant in the backward directions with respect to the pump, and we report the highest pump-to-Stokes conversion efficiency of 46 % for TE polarized light. Operating in the normal GVD regime using a resonator with higher $Q$ will enable a highly efficient Raman laser on the LN chip. Counter-propagating pump and Stokes fields might lead to richer nonlinear dynamics or functionalities such as symmetry breaking [49], counter-propagating solitons and Stokes solitons [50]. In addition, we observe nontrivial interactions between the Raman effect and $\chi^{(3)}$-based

FWM processes for Kerr comb formation. While the strong contribution from the Raman effect impedes soliton formation for TE polarization, we demonstrate soliton modelocking for TM polarization through optimization of the cavity geometry. TE-polarized phase-locked combs could alternatively be achieved with the help of strong electrical driving [5,41]. This work provides a deep insight on the dynamics and effects of Raman in the LNOI platform which critical for the design and development of chip-based nonlinear photonic devices.

## Materials and Methods

**Device parameters and fabrication**

For Raman characterization, we fabricate a racetrack microresonator from a commercial X-cut lithium niobate (LN) on insulator wafer with a thin-film LN thickness of 800 nm. The device is cladded with silicon oxide of 750-nm thickness with a top waveguide width of 1.2 μm and a slab thickness of 350 nm. The bending structure is based on Euler curves to avoid mode conversion between transverse electric (TE) and transverse magnetic (TM) modes. The two straight sections are each of 1.75 μm length along y-axis. The coupling gap between the bus waveguide and microresonator is 0.75 μm which leads to near critical coupling for TE modes and 45% transmission on resonance for TM modes at 1550 nm (see Supplementary Information). The FSR is 30 GHz. For Kerr comb generation, the microring resonator is fabricated on a 600-nm thickness thin-film LN wafer with a radius of 80 μm. The device is air-cladded with a slab thickness of 250 nm and a top waveguide width of 1.3 μm. The coupling gap is 830 nm which results in 50% transmission on resonance for TE mode and 83% for TM mode at 1550 nm.

Electro-beam lithography (EBL, 125 keV) is used for patterning the waveguides and microcavities in hydrogen silsesquioxane resist (FOX). Then the patterned LN wafer is etched using Ar$^+$-based reactive ion etching by a user-defined etch depth. The SiO$_2$ cladding is deposited by plasma-enhanced chemical vapor deposition. At last, the chip facet is diced and polished which typically results in a fiber-to-chip facet coupling loss of 7 dB.

**Comb characterization**

The group velocity dispersion (GVD) is simulated using a commercial finite element analysis software (COMSOL) based on the fabricated device geometry. Anomalous GVD is achieved for both TE and TM modes at telecommunication wavelengths (see Supplementary Information). A continuous-wave pump laser (Santec TSL-510) at 1570 nm is amplified by a L-band erbium doped fiber amplifier (EDFA, Manlight) and sent to the microring resonator using a lensed fiber after a polarization controller. The tuning of the laser is controlled by a piezo controller. The output is collected by an aspheric objective followed by a fiber collimator. A 90:10 fiber beamsplitter is used to separate the output light into two arms. The 10% arm is sent to an optical spectrum analyzer and the 90% arm is sent through a home-built 4-*f* shaper (effectively bandpass filter, 1575-1630 nm) to filter the pump. The filtered comb spectrum is sent to a photodetector (Newport 1811, 125 MHz) and a real-time spectrum analyzer (Tektronix RSA5126A). For Fig. 5c, a functional generator with a triangular function is sent to the piezo controller to scan the laser wavelength at 70 Hz, and the filtered comb is collected by a photodetector (Thorlabs, PDA05CF2) followed by an oscilloscope (Tektronix DPO2024, 200 MHz).

**Data Availability**

The data that support the plots within this paper and other finding of this study are available from the corresponding author upon reasonable request.

## Acknowledgments

Device fabrication is performed at the Harvard University Center for Nanoscale Systems (CNS), a member of the National Nanotechnology Coordinated Infrastructure Network (NNCI), which is supported by the National Science Foundation under NSF ECCS award no.1541959. This work was supported by the National Science Foundation (NSF) (ECCS-1740296 E2CDA), Defense Advanced Research Projects Agency (DARPA) (W31P4Q-15-1-0013), and Air Force Office of Scientific Research (AFOSR) (FA9550-15-1-0303).

## Conflict of Interests

The authors declare no competing financial interest.

## Contributions

M.Y. and Y.O. prepared the manuscript in discussion with all authors. M.Y. and Y.O. designed and performed the experiments. M. Y., Y. O., C. W., and M. Z. analyzed the data. R. C. and C. W. fabricated the devices. M.L. and A.L.G. supervised the project.

M.Y. and Y.O. contributed equally to this work.

1) **Figure Legends**

**Figure 1** | (a) Schematic of the LN crystalline structure. The crystal axis is along z axis. (b,c) Orientation of LN microresonator on an X-cut wafer. For Raman characterization a 30-GHz free spectral range (FSR) racetrack resonator is used (b). For Raman-Kerr interactions, a 250-GHz FSR microring resonator is used (c). TE and TM polarizations are also indicated. (d) Setup for characterization of forward (FW) and backward (BWD) Raman. EDFA: erbium doped fiber amplifier; BPF: band-pass filter; FPC: fiber polarization controller; OSA: optical spectrum analyzer.

**Figure 2** | Raman emission for TE polarized light from an X-cut LN racetrack microresonator. (a) Three Raman spectra (from top to bottom) at pump powers of 20, 200 and 400 mW, corresponding to the threshold power for the Raman oscillation with frequency shifts of 250 $cm^{-1}$, 628 $cm^{-1}$ and 875 $cm^{-1}$. The Stokes and pump are TE polarized and are along the LN polar z-axis. Both forward (blue) and backward (red) spectra are shown. The cascaded Raman peak from the 1$^{st}$ Stokes field is also observed at 1691 nm. (b) Raman emission power (top to bottom: 1$^{st}$ to 3$^{rd}$ Stokes) as a function of pump power. The threshold for the 1$^{st}$ Stokes in the backward direction is 20 mW with a high differential conversion efficiency of 46 % (linear fit shown with black dashed line). The deviation from the linear growth above 60mW of pump power is due to mini-combs formation and other Raman oscillations.

**Figure 3** | Raman emission for the TM mode (light polarization is along x-axis) at a frequency shift of 238 $cm^{-1}$ at the threshold power of 340 mW.

**Figure 4** | (a) Raman and Kerr oscillation for the TE mode from an LN microresonator as the pump laser is tuned into the cavity resonance (top to bottom). (b) RF spectrum corresponding to bottom spectra in (a).

**Figure 5** | (a) Kerr oscillation for the TM mode from an X-cut LN microresonator as the pump laser is red-detuned into the cavity resonance (top to bottom). (b) RF spectra corresponding to state (ii) and (iii) in (a). (c) Transmission measurement of the microresonator output beyond 1570 nm which excludes the pump wavelength. The laser is tuning to the resonance as the wavelength increases, dominated by the thermal-optic effect. (d) Zoom-in of the transmission trace in (c) which shows the onset of the soliton step.

2) **Tables**
    N. A.

3) **Figures**

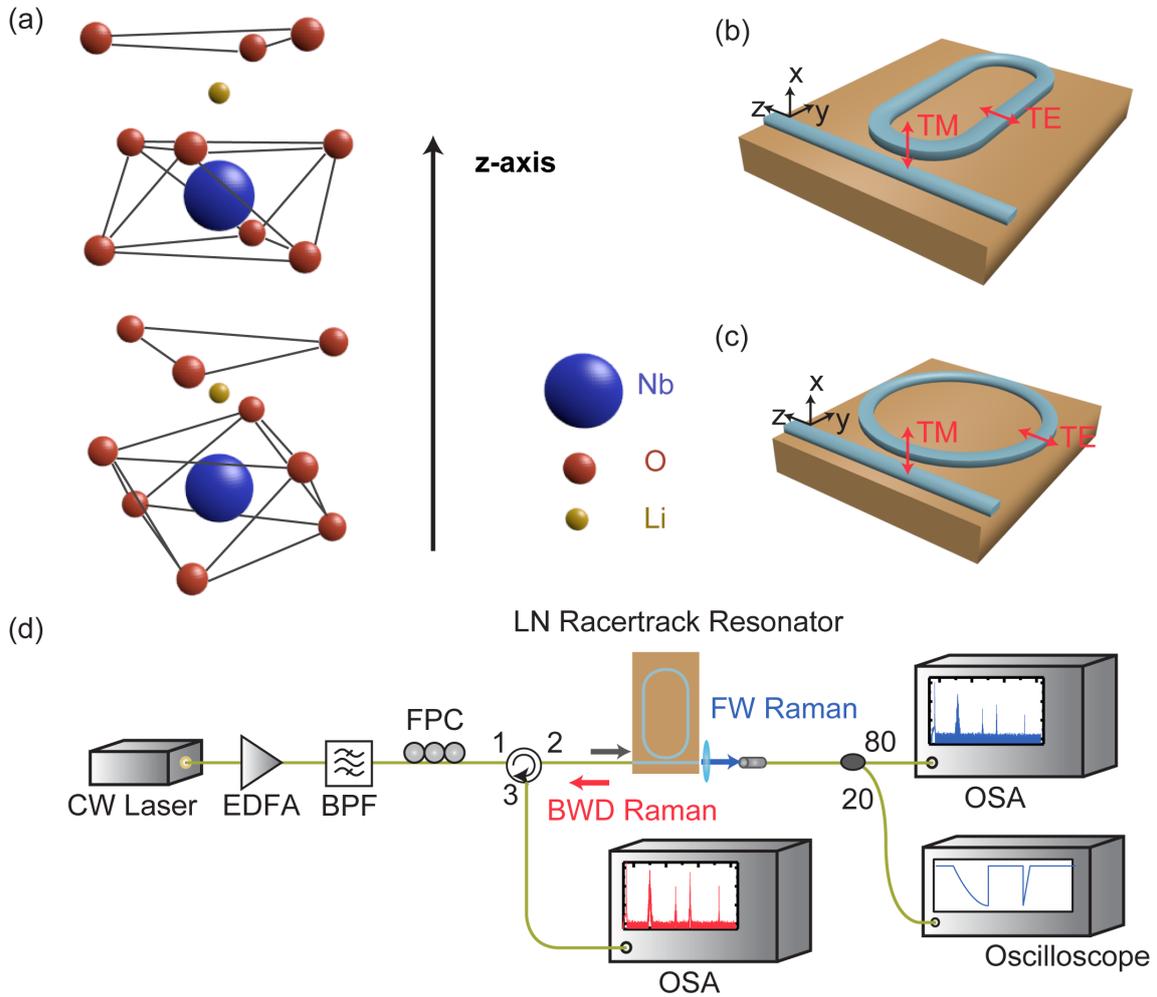

**Figure 1**

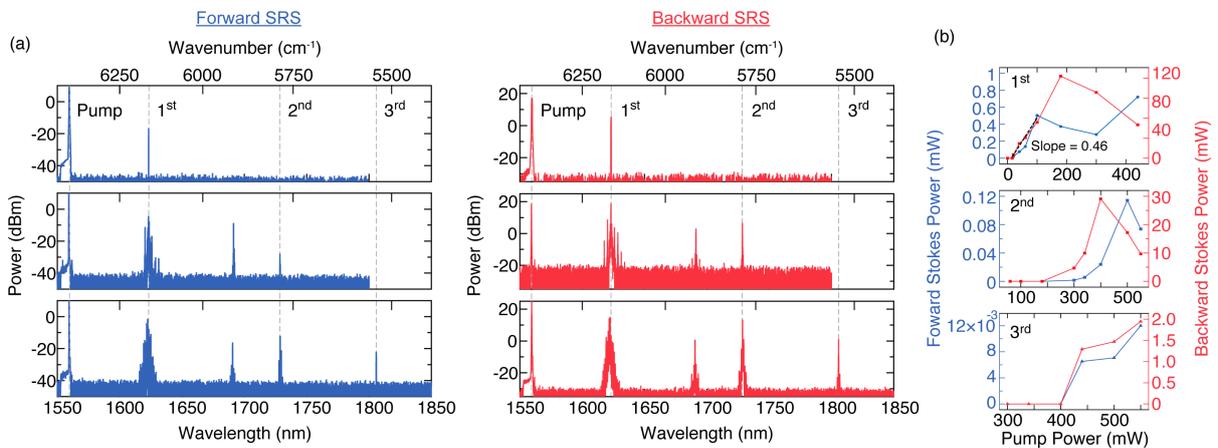

**Figure 2**

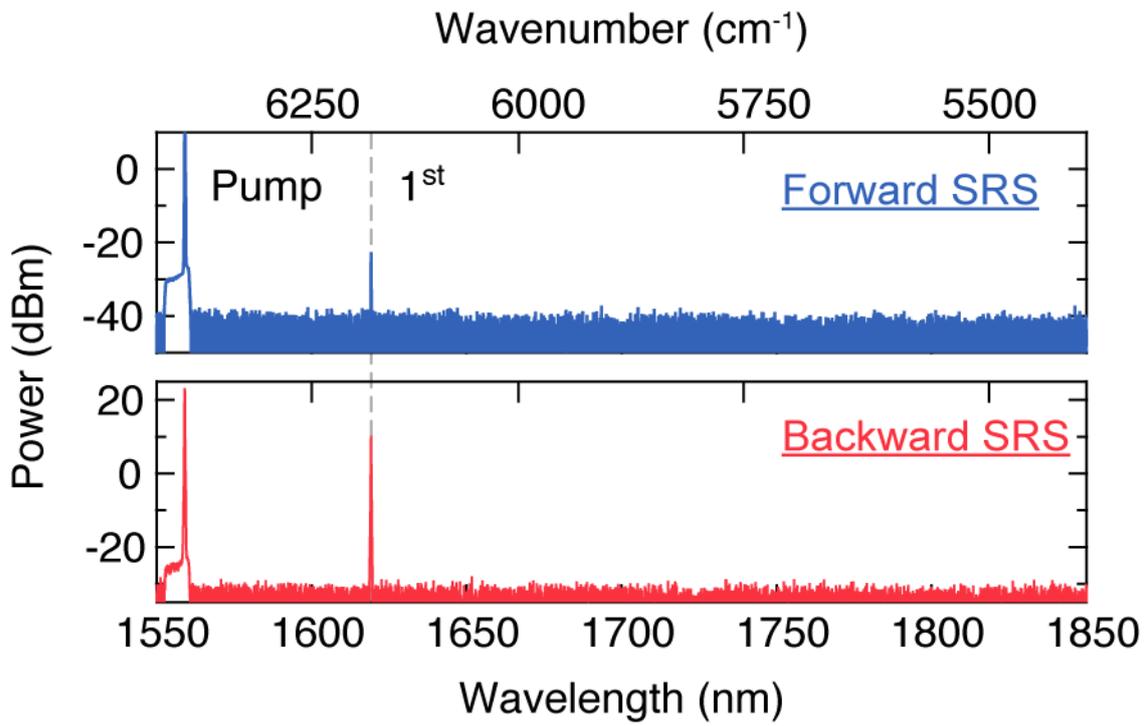

**Figure 3**

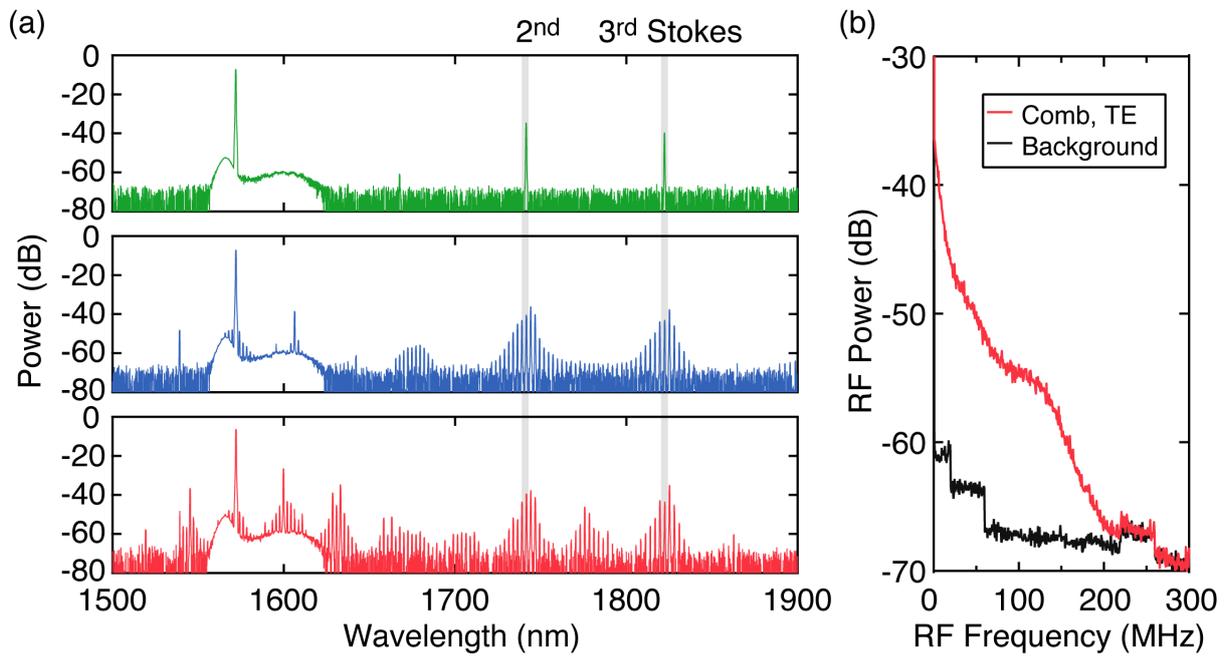

**Figure 4**

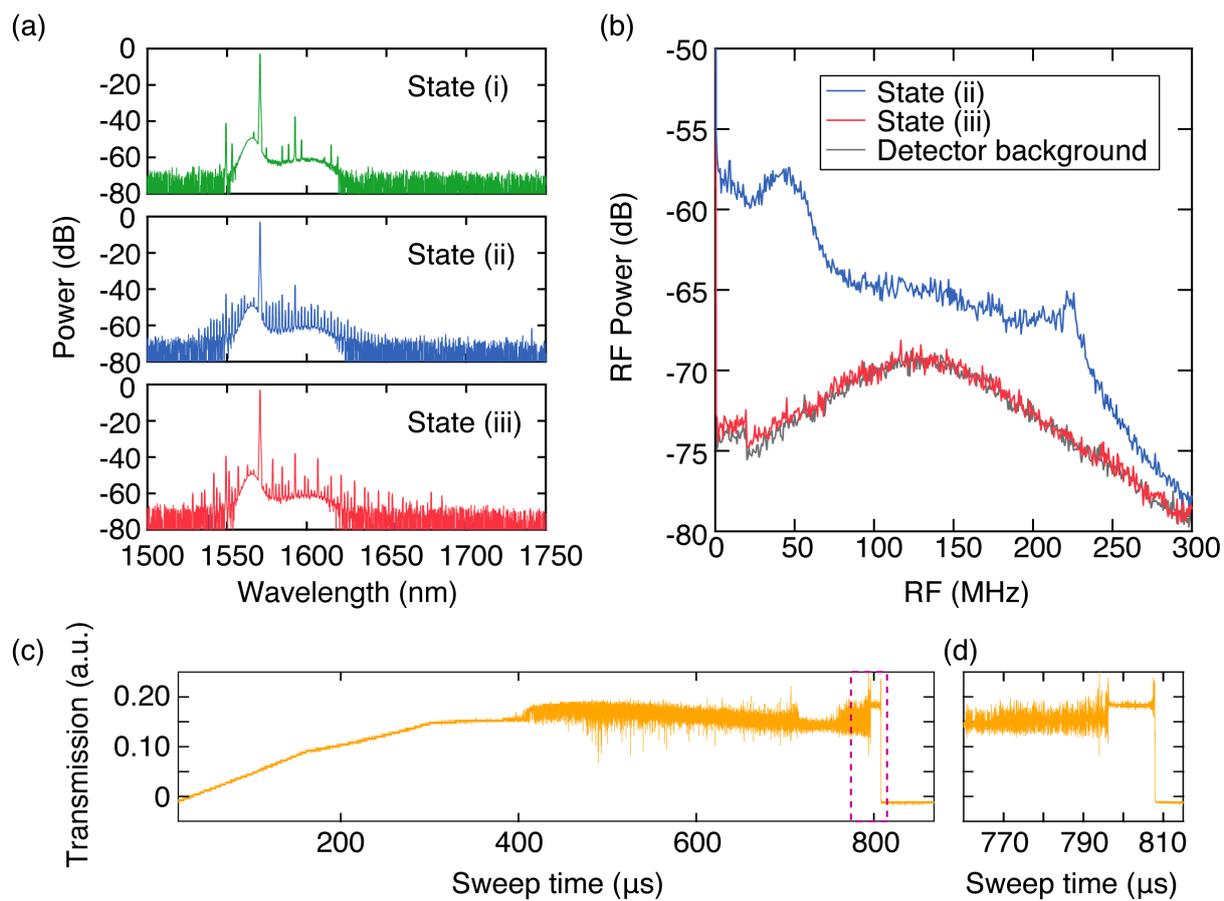

**Figure 5**